\documentclass[prb,aps,twocolumn,superscriptaddress]{revtex4-1}
\usepackage{bm}
\usepackage[colorlinks=true,linkcolor=blue,citecolor=blue]{hyperref}
\usepackage{times}
\usepackage{amsmath}
\usepackage{amssymb}
\usepackage{amsthm}
\usepackage{amsfonts}
\usepackage{enumerate}
\usepackage{latexsym}
\usepackage{ifpdf}
\newcommand{\beq}{\begin{equation}}
\newcommand{\eeq}{\end{equation}}
\usepackage{graphicx}
\usepackage{makeidx}
\usepackage{color}

\begin{document}

\title{Spin liquid behavior of a three-dimensional magnetic system Ba$_3$NiIr$_2$O$_9$ with $S$ = 1}
\author {Siddharth Kumar}
\affiliation{Department of Physics, Indian Institute of Science, Bengaluru 560012, India}
\author {S. K. Panda}
\affiliation{Department of Physics, Bennett University, Greater Noida 201310, Uttar Pradesh, India}
\author{Manju Mishra Patidar}
\affiliation{UGC-DAE Consortium for Scientific Research, University Campus, Khandwa Road, Indore 452017, India}
\author {Shashank Kumar Ojha}
\affiliation{Department of Physics, Indian Institute of Science, Bengaluru 560012, India}
\author {Prithwijit Mandal}
\affiliation{Department of Physics, Indian Institute of Science, Bengaluru 560012, India}
\author{Gangadhar Das}
\affiliation{Chemistry and Physics of Materials Unit, Jawaharlal Nehru Centre for Advanced Scientific Research, Jakkur, Bengaluru, 560064 India}
\author {J. W. Freeland}
\affiliation{Advanced Photon Source, Argonne National Laboratory, Argonne, Illinois 60439, USA}
\author{V. Ganesan}
\affiliation{UGC-DAE Consortium for Scientific Research, University Campus, Khandwa Road, Indore 452017, India}
\author {Peter J. Baker}
\affiliation{ISIS Pulsed Neutron and Muon Source, STFC Rutherford Appleton Laboratory, Harwell Campus, Didcot OX11 0QX, United Kingdom}
\author {S. Middey}
\email{smiddey@iisc.ac.in }
\affiliation{Department of Physics, Indian Institute of Science, Bengaluru 560012, India}

\begin{abstract}
The quantum spin liquid (QSL) is an exotic phase of magnetic materials where the spins continue to  fluctuate without any  symmetry breaking down to zero temperature.  Among the handful reports of QSL with   spin $S\ge$1, examples with magnetic ions  on a  three-dimensional magnetic lattice are extremely rare since both larger spin and higher dimension tend to suppress quantum fluctuations.  In this work, we offer a new strategy to achieve 3-D QSL with high spin by utilizing two types of transition metal  ions, both are magnetically active but located at crystallographically inequivalent positions. We design a 3-D magnetic system Ba$_3$NiIr$_2$O$_9$ consisting of interconnected corner shared NiO$_6$ octahedra and face shared Ir$_2$O$_9$ dimer, both having  triangular arrangements in \textit{a-b} plane. X-ray absorption spectroscopy measurements  confirm the presence of Ni$^{2+}$ ($S$=1). Our detailed thermodynamic and magnetic measurements reveal that this compound is a realization of gapless QSL state down to at least 100 mK. Ab-initio calculations find a strong magnetic exchange between Ir and Ni sublattices and  in-plane antiferromagnetic  coupling between the dimers, resulting in dynamically fluctuating magnetic moments  on both the Ir and Ni sublattice.

 \end{abstract}

\maketitle
	
Experimental realization and theoretical description of the highly entangled quantum spin liquid phase remain challenging topics of quantum many-body physics~\cite{anderson:1973p153}.
Over the last fifteen years, several spin-1/2 systems with two-dimensional frustrated lattice  have been reported as probable candidates with QSL behavior~\cite{balents:2010p199,savary:2016p016502,zhou:2017p025003,knolle:2019p451,broholm:2020p367}. Since experimental reports of QSL with either spin $(S) \ge$ 1 ~\cite{Cheng2011,Chamorro:2018p034404} or three dimensional arrangement of spins~\cite{okamoto:2007p137207,Koteswararao:2014p035141,balz:2016p942,gao:2019p1052,chillal:2020p1} are very few, it can be easily anticipated that  chances of having a QSL with both higher spin and 3-D geometry is extremely low~\cite{plumb2019:2019p54}. Many of the attempts to obtain spin-1 QSL have been  focused on stabilizing Ni within various structural network~\cite{nakatsuji:2005p1697,Cheng2011, lu:2018p094412,Chamorro:2018p034404,plumb2019:2019p54,medrano:2018p054435}.
However, the magnetic behavior of these $S$ = 1 systems  at low temperature differs widely from each other, even in the compounds with similar structural geometry. For example,  unlike to the well known 120$^\circ$ spin structure~\cite{starykh:2015p052502}, as  observed in $A_3$NiNb$_2$O$_9$~\cite{lu:2018p094412},  Ba$_3$NiSb$_2$O$_9$ shows characteristic spin liquid behavior~\cite{Cheng2011,Fak2017,Quilliam2016}, whereas NiGa$_2$S$_4$ hosts a spin nematic phase~\cite{Bhattacharjee:2006p092406}. The interaction  of such Ni-based $S$ = 1 triangular lattice with another magnetically active sublattice might result in an exotic magnetic phase in three-dimensional materials. However, only very few 3-D compounds with such feasibility exist  ~\cite{treiber:1982p189,lightfoot:1990p174,ferreira:2018p2973}.

\begin{figure*}
		\centering
		\includegraphics[width=\linewidth]{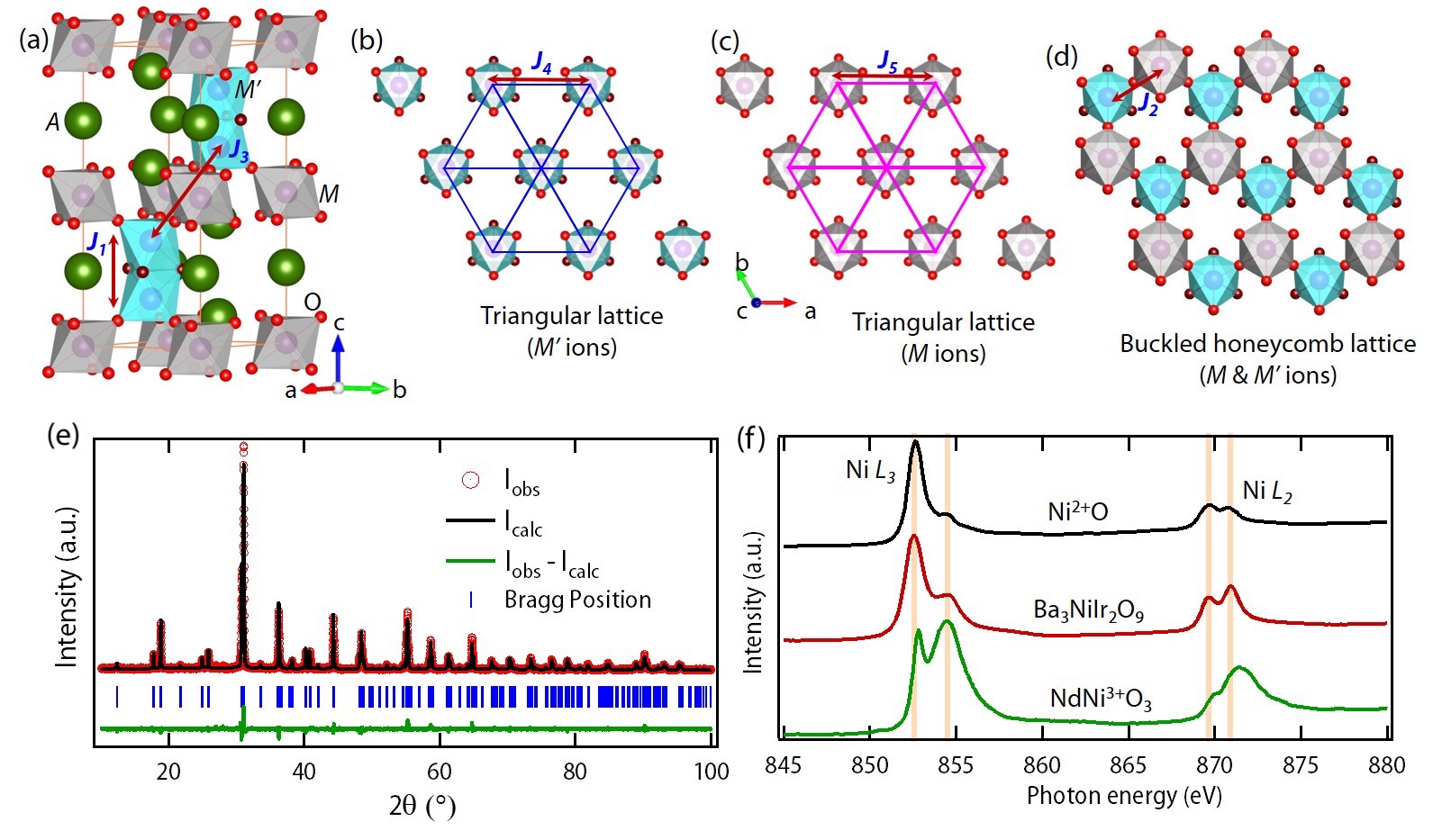}
		\caption{\label{Fig.1} (a) Unit cell of 6$H$ $A_3MM^\prime$O$_9$ without any disorder. (b), \& (c) shows the triangular lattice arrangement of $M^\prime_2$O$_9$ dimers, $M$O$_6$ octahedra, respectively, in \textit{a-b} plane.  (d) $M$-O-$M\prime$ connectivity forms buckled honeycomb lattice. $A$, $M$ and $M^\prime$ corresponds to Ba, Ni, and Ir, respectively, for BNIO. Different magnetic exchange pathways ($J_i$) are also shown in (a)-(d). (e) Observed and refined powder XRD pattern of BNIO. (f) XAS spectrum of Ni $L_{3,2}$-edge of BNIO  along with NiO  and NdNiO$_3$  for comparison. The XAS data for NdNiO$_3$ has been adapted from Ref.~\onlinecite{freeland:2016p56}.}
	\end{figure*}

Six-layered hexagonal (6$H$)  perovskite $A_3MM^\prime_2$O$_9$ (Fig.~\ref{Fig.1}(a)) with magnetic $M$ and $M^\prime$ ions constitutes a 3-D spin system. Both, $M$O$_6$ octahedral units and face-shared $M^\prime_2$O$_9$ dimers form a triangular lattice in $a$-$b$ plane (Fig.~\ref{Fig.1}(b)-(c)) and would become geometrically frustrated in the presence of antiferromagnetic interaction. Moreover, the $M$-O-$M^\prime$ connectivity  constitutes a buckled honeycomb lattice (Fig.~\ref{Fig.1}(d)), which could host a Kitaev spin liquid phase in case of spin-1/2 ions with strong spin-orbit coupling (SOC)~\cite{kitaev:2006p2,takagi:2019p264}. In search of SOC driven elusive nonmagnetic $J$ = 0 state and excitonic magnetism in $d^4$ system, several Ba$_3M$Ir$_2$O$_9$ compounds with nonmagnetic $M^{2+}$ have been investigated recently~\cite{Khaliullin2013,Nag2016,Nag2018,Khan:2019p064423}. However,  the comparable strength of  SOC, non-cubic crystal field, Hund's coupling, and superexchange interaction gives rise to a small but finite moment on the Ir sublattice. Moreover, interdimer hopping   results in a spin-orbital liquid phase in  Ba$_3$ZnIr$_2$O$_9$~\cite{Nag2016}. Replacing the nonmagnetic Zn$^{2+}$ by an isovalent magnetic  ion such as Ni$^{2+}$ should provide a unique opportunity to investigate the magnetic response of a triangular lattice with $S$ = 1 in presence of the interconnected Ir sublattice with  fluctuating magnetic moments. If both Ni and Ir moments of Ba$_3$NiIr$_2$O$_9$  fluctuate dynamically, then it would offer a new route to realize 3-D QSL by involving two different magnetic ions. Also, it would be  a 3-D QSL with a new type of structural network as compared to all existing examples with pyrochlore~\cite{gao:2019p1052,plumb2019:2019p54}, hyperkagome~\cite{okamoto:2007p137207}, and hyper hyperkagome structure~\cite{Koteswararao:2014p035141,chillal:2020p1}.

 In this paper, we report on the electronic and magnetic behavior of Ba$_3$NiIr$_2$O$_9$ (BNIO). The phase purity and absence of any cationic site disorder have been demonstrated by powder X-ray diffraction (XRD) measurement. X-ray absorption spectroscopy (XAS) experiments have confirmed the desired +2 oxidation state of Ni. Persisting  spin fluctuations down to 100 mK have been revealed by magnetization, specific heat and muon spin-rotation ($\mu$SR) measurements. We have also investigated  BNIO by density functional theory calculations including Hubbard $U$ and SOC within the framework of LSDA+$U$ (local spin density approximation + $U$) approach.  We have found  not only appreciable magnetic exchange between Ni and Ir-sublattice but also  antiferromagnetic coupling in the triangular sublattice of Ir. This geometrical frustration  prohibits any long-range magnetic ordering and makes BNIO a rare example of three-dimensional QSL involving $S$ = 1.

\section*{Results}
 Polycrystalline  BNIO was synthesized by solid state synthesis route.  Members of the $A_3MM^\prime_2$O$_9$ series can have site disorder between face shared, and corner shared octahedra~\cite{Middey:2011p144419}. It is also well known that structural disorder  often jeopardizes QSL behavior, resulting in magnetic order or spin glass freezing~\cite{zhong:2019p14505}.   All peaks of the powder XRD  pattern of BNIO (Fig.~\ref{Fig.1}(e)) can be indexed and refined with 6$H$ structure having space group $P6_3/mmc$.  The refinement also confirms that all corner-shared (face-shared) octahderal units  are occupied by Ni (Ir) without any Ni-Ir site disorder. The  structural  parameters obtained from the refinement have been listed in SM~\cite{sup}. The temperature dependent XRD measurements down to 15 K also rules out any structural transition. Having confirmed that both Ni and Ir ions form triangular lattices in \textit{a-b} plane without any disorder, we want to verify whether Ni has  indeed $S$ =  1 state. For this purpose, XAS measurements were carried out~\cite{freeland:2016p56}.   The comparison of Ni  $L_{3,2}$ XAS line shape and  energy  position  of  BNIO, Ni$^{2+}$O, and NdNi$^{3+}$O$_3$ (Fig. ~\ref{Fig.1}(e)) testifies the desired +2 oxidation of Ni in present case. The octahedral crystal field of Ni$^{2+}$ ($d^8$: $t_{2g}^6$, $e_g^2$) ensures $S$ = 1 on Ni sublattice.

\begin{figure*}
	\centering
	\includegraphics[width=\linewidth]{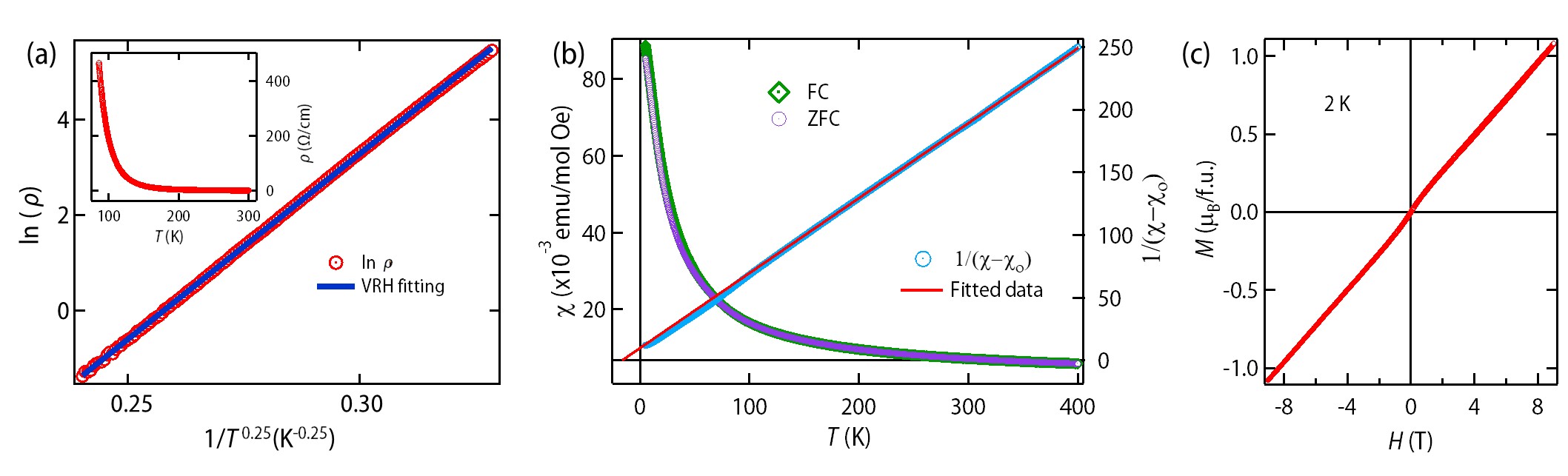}
	\caption{\label{Fig.2} (a)  Fitting of $\rho$ by 3-D variable range hopping model for BNIO. Inset shows $\rho$ vs. $T$. (b) $\chi$ vs\textit{ T } on left axis and $1/(\chi-\chi_0)$ along with fitting on right axis. (c) \textit{M-H} at 2 K. }
\end{figure*}

Electrical measurement demonstrates insulating nature of the sample (inset of Fig.\ref{Fig.2} (a)), which can be fitted using Mott's variable range hopping (VRH)~\cite{Hill1976}  model in three-dimensions ($\rho = \rho_o\exp(T_o/T)^{1/4}$) as shown in Fig.\ref{Fig.2} (a). We also note that the insulating behavior of Ba$_3$ZnIr$_2$O$_9$ is well described by VRH in two-dimensions. This difference between two compounds is a manifestation of electron hopping paths along the Ni-O-Ir bonds. Our electronic structure calculations (see SM~\cite{sup}) further demonstrate  that the insulating state can be obtained only by considering both electron correlation and SOC, implying that BNIO is a SOC driven Mott insulator~\cite{Kim:2008p076402}.

The temperature dependent field cooled and zero-field cooled  magnetic susceptibility does not differ from each other and also do not show any anomaly (Fig.~\ref{Fig.2}(b)). This strongly implies absence of any long range magnetic ordering and spin glass behavior.
We have fitted the data by a modified Cuire-Weiss law ($\chi$ = $\chi_0$ + $\frac{C_W}{T-\theta_{CW}}$) where $\chi_0$, $C_W$ and $\theta_{CW}$ represents temperature independent susceptibility contribution, Curie constant and Curie temperature,  respectively. The fitting, shown as plot of 1/($\chi$-$\chi_0$) vs. $T$ in right axis of Fig.~\ref{Fig.2}(b), results a $\theta_{CW}\sim$  -15 K. Negative values of Curie-Weiss temperature signify net antiferromagnetic interaction among the spins of BNIO. The relatively smaller value of $\theta_{CW}$ is related to the presence of multiple  exchange pathways, which will be discussed   in later part of this paper.  The effective magnetic moment  $\mu_{eff}$ (= $\sqrt{8C_W}$)  is found to be $\sim$ 3.65 $\mu_B$ from the fitting.  Interestingly, the effective magnetic moment of a similar compound Ba$_3$NiSb$_2$O$_9$, with nonmagnetic  Sb$^{5+}$ was reported to be around 3.54 $\mu_B$~\cite{Cheng2011}. This  gives an estimate of $g$-factor  $\sim$ 2.5, similar to other Ni$^{2+}$ based systems~\cite{carlin_paramagnetism_1986}. If we assume similar value for the present compound, the effective Ir moment turns out to be  0.9 $\mu_B$ per  dimer  (= $\sqrt{\mu_{eff}^2-\mu_{Ni}^2}$) i.e. 0.64 $\mu_B$/Ir. This  is very similar to the Ir moment (0.5 - 0.6 $\mu_B$) reported for  Ba$_3$MgIr$_2$O$_9$~\cite{Nag2018}, though a nonmagnetic $J$ = 0 state is expected for Ir$^{5+}$ from a pure ionic picture. Thus, our analysis highlights both Ni$^{2+}$ and Ir$^{5+}$  participate in magnetism of BNIO compound.   The estimated magnetic moments from our LSDA+U+SOC calculations (shown in SM~\cite{sup})  are also in good agreement with our experimental results. Fig.~\ref{Fig.2} (c) shows $M$-$H$ done at 2 K between $\pm$ 9 T. The absence of any hysteresis again confirms absence of ferromagnetism and spin glass freezing at 2 K. The presence of antiferromagnetic interaction without any long range magnetic ordering or spin glass transition strongly indicates that BNIO is  a favorable candidate of QSL.

\begin{figure*}
	\centering
	\includegraphics[scale=0.7]{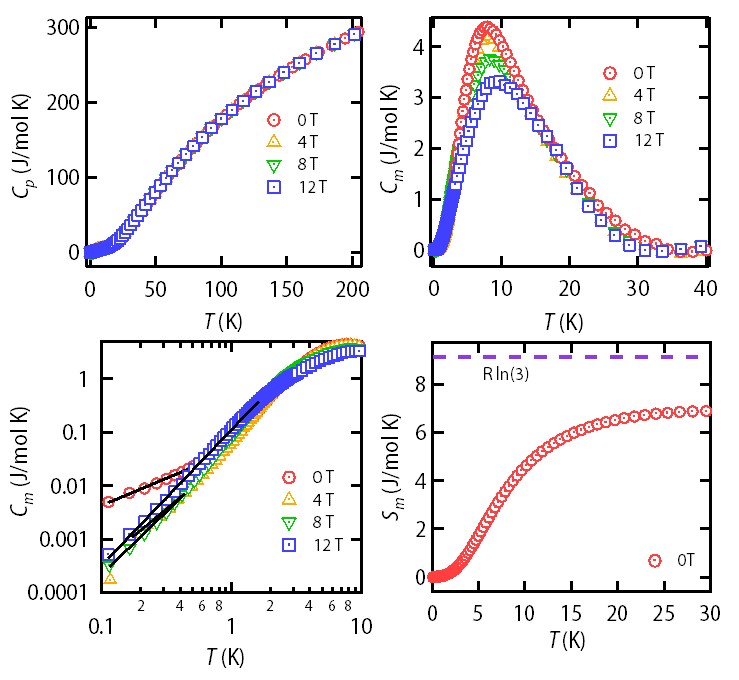}
	\caption{\label{Fig.3}(a) $C_p$ vs \textit{T} curves at various fields. (b) Magnetic specific heat ($C_m$) extracted by subtracting lattice contribution. (c) Low temperature part of $C_m$ plotted in log-log plot. (d) Magnetic entropy for zero field.}
\end{figure*}

In order to further investigate the magnetic nature of the sample,  we have measured specific heat ($C_p$) from 100 mK to 200 K.  $C_p$ not only  probes the presence/absence of any long-range magnetic ordering but also  provides very crucial information about the nature of low energy excitation. The absence of any $\lambda$-anomaly (Fig.~\ref{Fig.3}(a)) again confirms  absence of long-range order  and/or  any structural transition down to 100 mK, consistent with the magnetic measurement and XRD results, respectively. For an insulating compound with magnetic spins,  $C_p$ consists of  lattice specific heat ($C_{lat}$) and   magnetic specific heat ($C_m$).  In absence of any analogous non-magnetic compound, the contribution of $C_{lat}$ has been evaluated by fitting $C_p$ in 30 K - 200 K range by a Debye-Einstein equation with one Debye term and two Einstein terms (details are in SM~\cite{sup}) and extrapolating  the fitted curve down to 100 mK. A broad peak is observed   around 7 K in $C_m$ vs. $T$ plot (Fig.~\ref{Fig.3}(a)). We can not capture this  feature by Schottky anomaly, arising due to energy level splitting (see SM~\cite{sup}). On the other hand, such feature is commonly observed in  spin liquid materials and thus, could be considered as a signature of crossover from thermally disordered paramagnet to quantum disordered spin liquid state~\cite{Okamoto2007,balents:2010p199,Li:2015p1,Dey2017}.  The position of this broad peak shows negligible shifts with the application of magnetic field (shifts $\sim$ 1 K for applied field of 12 T).

At low temperature, $C_m$ follows power-law behavior $C_m$ = $\gamma T^\alpha$ (Fig.~\ref{Fig.3}(c)). For zero field, the magnitude of coefficient $\gamma$ is 45 mJ/mol K$^2$ and the exponent $\alpha$ is 1.0$\pm$0.05 within 0.1 K - 0.6 K range. The value of $\gamma$ is very similar  to other gapless spin liquid candidates: like Ba$_3$CuSb$_2$O$_9$ (43.4 mJ/mol K$^2$)~\cite{Zhou2011}, Ba$_2$YIrO$_6$ (44 mJ/mol K$^2$)~\cite{NagAPRB2018}, Ba$_3$ZnIr$_2$O$_9$ (25.9 mJ/mol K$^2$)~\cite{Nag2016}, Sr$_2$Cu(Te$_{0.5}$W$_{0.5}$)O$_6$ (54.2 mJ/mol K$^2$)~\cite{Mustonen2018}. Also, the linear $T$ behavior with nonzero $\gamma$ in an insulator comes due to gapless spinon excitations with a Fermi surface and has been reported in several organic and inorganic spin liquid candidates~\cite{Yamashita2008,Yamashita2011,Zhou2011,Cheng2011,Mustonen2018,Clark2014,Uematsu2019PRL}. Otherwise, any gapped excitation would result in an exponential dependence of $C_m$ on $T$.   $\alpha$ becomes 2.6$\pm$0.05 within 0.8 K - 2.1 K. Interestingly, the application of an external field destroys the linear $T$ behavior of $C_m$. For $\mu_0H$ = 4 T, $\alpha$ becomes 2 for 0.15 K $\leq T \leq$ 0.50 K and 2.9 for 0.5 K $\leq T\leq$ 2.4 K. We note that $\alpha$ is found to be between 2 to 3 for several spin nematic phase~\cite{nakatsuji:2005p1697,NakatsujiPRL2007,Povarov2019,Kohama2019}. Further studies are necessary to investigate the possibility of transition from spin liquid to spin nematic phase by the application of a magnetic field.
The amount of released magnetic entropy ($S_m$)  is evaluated by integrating $C_m/T$ w.r.t. $T$ and is shown in Fig.~\ref{Fig.3}(d). For BNIO,  the entropy saturates at 6.9 J/mol K for  zero field measurement, which is only 75\% of the total entropy expected for even a $S = 1$ system [$R$ln(2$S$+1), where $R$ is universal gas constant]. The retention of a large amount of entropy at low temperature is another signature of the spin-liquid nature of BNIO, which has been reported as well for many other QSL~\cite{Zhou2011,Cheng2011,Mustonen2018,Yamashita2008}.

\begin{figure}
	\centering
	\includegraphics[width=\linewidth]{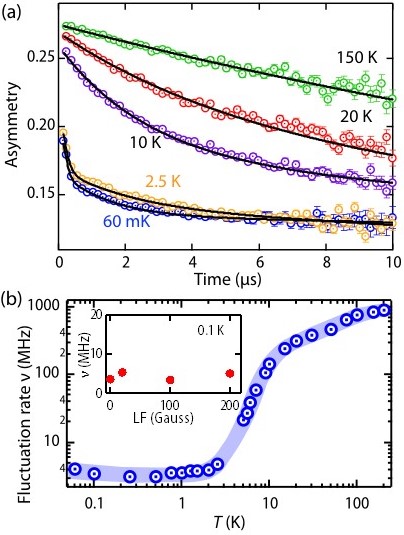}
	\caption{\label{Fig.4} (a) Asymmetry vs time curves at various temperatures taken at zero magnetic field and fitted curve (solid line) using equation 1.  (b) Variation of $\nu$ to temperature (shaded area is a guide to the eye). Inset shows variation of $\nu$ with applied LF at 100 mK}
\end{figure}

	To have a further understanding of the magnetic behavior at low temperature, we have performed $\mu$SR measurements, which is a very sensitive local probe to detect a weak magnetic field, even of the order of 0.1 Oe~\cite{Blundell1999}.  Fig.~\ref{Fig.4}(a) shows asymmetry vs time curves for zero-field (ZF) measurements at selected temperatures. Neither any oscillation nor the characteristic 1/3$^\mathrm{rd}$ recovery of the asymmetry is observed down to 60 mK, strongly implying the  absence of long-range magnetic  ordering or spin freezing. For a  system with two interacting spin networks, the local magnetic field, felt by a muon at a stopping site is contributed by both magnetic sublattices. In such cases, the muon relaxation function is generally described  by a product of two response functions, representing local fields from  two spin networks~\cite{Uemura:1985p546}. However, our such attempts  considering different possible combinations of relaxation functions, including a  spin glass like relaxation~\cite{Morenzoni:2008p147205,Uemura:1985p546}, did not provide a satisfactory fitting of the experimental observed data (see SM~\cite{sup}). This further supports the absence of spin glass freezing in present case.

Interestingly, similar to the other hexagonal Ba$_3M$Ir$_2$O$_9$~\cite{Nag2016,Nag2018}, these asymmetry curves consist of  one fast relaxing, one moderately relaxing, and one hardly relaxing components. We  have fitted these curves using a combination of two dynamical relaxation functions with a common fluctuation rate $\nu$ and a Kubo-Toyabe function (KT)~\cite{Hayano:1979p850},
\begin{equation}
	\label{eq_musr}
	A(t) = A_1G(t,\Delta H_1,\nu)+A_2G(t,\Delta H_2,\nu)+A_3KT(t,\delta)
\end{equation}
where $A_1$, $A_2$, $A_3$ are amplitudes.
The static KT function, corresponding to the hardly relaxing component, accounts for the muons stopping at the silver sample holder as we  find that the relaxation curve from the bare sample holder can also be described by a  KT function with similar $\delta$. The dynamical relaxation, arising due to the presence of a  field distribution ($\Delta H$) with a  fluctuation rate ($\nu$) is represented by the Keren function $G(t,\Delta H,\nu)$~\cite{Keren:1994p10039}. The presence of two dynamical relaxations implies  two inequivalent muon stopping sites, which are likely to be related with the two types of crystallographically inequivalent oxygen in hexagonal  Ba$_3M$Ir$_2$O$_9$~\cite{Nag2016,Nag2018}. The asymmetry data over a large temperature range (60 mK - 150 K) have been fitted by allowing $\nu$ to vary  with $T$ and, the extracted values of $\nu$ has been shown as a function of $T$ in Fig.~\ref{Fig.4}(b). The background contribution is different between measurements in dilution refrigerator and helium cryostat and,  has been kept fixed for our analysis within the corresponding temperature range.
The inequality $\nu> \gamma \Delta H$ ( $\gamma$ = muon gyromagnetic ratio =  2$\pi\times$135.5 Mrad s$^{-1}$ T$^{-1}$) holds for both relaxing components as $\gamma \Delta H_1\sim$0.425 MHz and $\gamma \Delta H_2\sim$ 0.09 MHz for the lowest  temperature of our measurement (60 mK). This justifies the use of dynamical relaxation functions and establishes spin liquid nature of BNIO. We note that the value of $\nu$  ($\sim$ 4 MHz) at low temperature for BNIO is one order of magnitude smaller than Ba$_3$ZnIr$_2$O$_9$~\cite{Nag2016} and is likely to be related  with the involvement of large spins on the  Ni sublattice.

$\mu$-SR spectra, recorded at 100 mK in presence of an applied longitudinal field (LF) have further corroborated  QSL nature of BNIO. In case of relaxation arising from a static internal field distribution with width $\Delta H_i$, an applied LF $\sim$ 5-10$\Delta H_i$ would completely  suppress the relaxation.  From the analysis of ZF  $\mu$SR data, we found $\Delta H_1 \sim$ 5 Gauss and $\Delta H_2 \sim$ 1 Gauss.  No such decoupling is observed in the present case in measurement up to 200 Gauss (see inset of Fig. 4(b) and SM~\cite{sup}),   establishing the dynamic nature of spins in BNIO down to atleast 100 mK.

\section*{Discussion}
To understand the underlying mechanism of the observed QSL state,   we estimated the inter-atomic magnetic exchange interactions from the converged LSDA+$U$+SOC calculations using the formalism of Ref.~\onlinecite{Kvashnin:2015p125133} (see Method section and SM~\cite{sup} for details). As shown in Table-I,  the strongest interaction is antiferromagnetic, which is between the Ir ions of the structural dimer. The strong Ir-Ni interaction further testifies three-dimensional nature of BNIO. Most importantly, Ir-Ir exchange in the \textit{a-b} plane ($J_4$) is found to be  antiferromagnetic, resulting in-plane magnetic frustration and explains the origin of the QSL behavior of the present system. However, the presence of strong Ir-Ni and Ni-Ni ferromagnetic exchange reduces the net antiferromagnetic exchange of this system, resulting a relatively lower value of negative $\theta_{CW}$. We further  note that the ferromagnetic Ni-Ni exchange has been  observed also in antiferromagnetic phase of analogous compound Ba$_3$NiRu$_2$O$_9$~\cite{lightfoot:1990p174}.

  	\begin{table}[h]
	\caption{\label{T1}Exchange couplings obtained from $ab$-initio calculations. Exchange pathways have been shown in Fig.~\ref{Fig.1}. AFM and FM refers to antiferromagnetic and ferromagnetic interaction, respectively. }
	\centering
	\begin{tabular}{lccccc}
		\hline
		\hline
		Exchange & Interacting  & Number & Magnitude & Type& $|$z$_iJ_i$/$J_1|$\\
		($J_i$)& pair           &  of neighbor ($z_i$) & (meV) & & \\
		\hline
		$J_1$ &	Ir-Ir &  1 & -8.91 & AFM& 1\\
		$J_2$ & Ir-Ni & 3 & 0.96 & FM &0.32 \\
		$J_3$ & Ir-Ir & 3 & 0.10 & FM &0.03 \\
		$J_4$ & Ir-Ir & 6 & -0.17 & AFM &0.11 \\
		$J_5$ & Ni-Ni & 6 & 0.09 & FM &0.06 \\
		\hline
		\hline
	\end{tabular}
\end{table}

To summarize, our detailed measurements reveal that 6$H$ BNIO containing $S$=1 hosts a gapless spin liquid phase below 2 K.  The involvement of Ir$^{5+}$ and Ni$^{2+}$  in the magnetic properties of BNIO is revealed by dc magnetic measurements, $\mu$SR experiments, and electronic structure calculation. The antiferromagnetic  interaction between Ir$_2$O$_9$ dimers in \textit{a-b} plane facilitates geometrical frustration driven QSL phase of BNIO. Since many Ba$_3$$MM_2^\prime$O9 compounds can be stabilized in 6$H$ structure, we believe this work will lead to the realization of many 3-D QSLs with large spin by judicial choice of $M$ and $M^\prime$.

\subsection*{Method}

Stoichiometric amount of BaCO$_3$, NiO and Ir metal power were used as starting materials for the solid state synthesis of BNIO.  The mixture was heated multiple times at 1175$^\circ$\ C with intermediate grindings till the desired phase is formed.  Powder XRD was carried out using a lab based Rigaku Smartlab diffractometer and also in the Indian beamline (BL-18B) at the Photon Factory, KEK, Japan. The diffraction pattern of the final phase was refined by Reitveild method using FULLPROF~\cite{Rodriguez-Carvajal1993}.

 XAS spectra of Ni $L_{3,2}$-edges were recorded in bulk sensitive total fluorescence yield mode in 4-ID-C beam line of Advanced Photon Source, USA.  DC magnetic measurements were carried using a Quantum Design (QD) SQUID magnetometer. Heat capacity measurements ($C_p$) were done  in a dilution refrigerator insert coupled with a 16T QD-PPMS system using relaxation calorimetry.  $\mu$SR experiments down to 60 mK were performed using pulsed muon beam at MuSR spectrometer of ISIS Neutron and Muon Source, UK. A dilution fridge was used to record $\mu$SR data from 60 mK to 4 K and a cryostat was used for temperatures above 1.5 K.

The density functional theory (DFT) calculations have been performed in the local spin-density approximation + Hubbard $U$ (LSDA+U) approach with and without including spin-orbit coupling (SOC) by means of a full potential linearized muffin-tin orbital method (FP-LMTO)~\cite{FPLMTO_Orig,FPLMTO} as implemented in the RSPt code~\cite{FPLMTOCode}.
The Brillouin-zone (BZ) integration is carried out by using the thermal smearing method with 10 $\times$ 10 $\times$ 4 k-mesh.
For the charge density and potential angular decomposition inside the muffin-tin (MT) spheres, the value of maximum angular momentum was taken equal to $l_{max} = 8$. To describe the electron-electron correlation within LSDA+U approach, we have taken {$U$ = 6 eV}, {$J$ =0.8 eV} for Ni-$d$ states and {$U$ = 2 eV}, {$J$ =0.6 eV} for the Ir-$d$ states.
The set of the correlated orbitals located on Ni and Ir sites were obtained by projecting the electron density onto the corresponding MT sphere with a certain angular character (so-called ``MT-heads" projection\cite{grechnev-FeCoNi-PRB}).
\par
After obtaining the self-consistent fully converged LSDA+$U$+SOC calculations, the magnetic force theorem~\cite{lichtenstein-exch, PhysRevB.61.8906} was used to extract the effective inter-site magnetic-interaction parameters ($J_{ij}$). In this approach the magnetic system is mapped onto the Heisenberg Hamiltonian:
\begin{equation}
\hat H = - \sum_{i \neq j} J_{ij} \vec{S_i} \cdot \vec{S_j} .
\label{HH}
\end{equation}
Further, $J_{ij}$ are extracted in a linear-response manner via Green's function technique. A detailed discussion of the
implementation of the magnetic force theorem in RSPt is provided in Ref.~\onlinecite{PhysRevB.91.125133}. This method is considered to be one of the most accurate techniques for the estimation of exchange interactions and also been successfully employed for many transition metal compounds~\cite{PhysRevB.94.064427}.

\subsection*{Acknowledgement}
S.M. acknowledges financial support from ISRO-IISc Space Technology Cell and Infosys Foundation, Bangalore. S.M. and S.K. acknowledge insightful discussions with Dr. Subhro Bhattacharjee, Dr. Yogesh Singh, Dr. Pabitra Biswas.  Authors thank the Department of Science and Technology, India for the financial support, Saha Institute of Nuclear Physics and Jawaharlal Nehru Centre for Advanced Scientific Research, India for facilitating the experiments at the Indian Beamline, Photon Factory, KEK, Japan. The authors also acknowledge the experimental support from UGC-DAE CSR, Indore, staffs of LT \& Cryogenics especially Er. P. Saravanan for their technical support. This research used resources of	the Advanced Photon Source, a U.S. Department of Energy Office of Science User Facility operated by Argonne National Laboratory under Contract No. DE-AC02-06CH11357. Experiments at the ISIS Neutron and Muon Source were supported by a beamtime allocation RB1920396 from the Science and Technology Facilities Council.

%

\end{document}